# A Secure and Efficient Approach for Issuing KYC Token As COVID-19 Health Certificate Based on Stellar Blockchain Network

Kiarash Shamsi, Koosha Esmaeilzadeh Khorasani, Mohammad Javad Shayegan*
Department of Computer Engineering, University of Science and Culture, Tehran, Iran
kiarash.shamsi@gmail.com, k.e.khorasani@gmail.com, shayegan@usc.ac.ir

*Abstract*— Today's world is struggling with the COVID-19 pandemic, as one of the greatest challenges of the 21st century. During the lockdown caused by this disease, many financial losses have been inflicted on people and all industries. One of the fastest ways to save these industries from the COVID-19 or any possible pandemic in the future is to provide a reliable, fast, smart, and secure solution for people's health assessment. In this article, blockchain technology is used to propose a model which provides and validates the health certificates for people who travel or present in society. For this purpose, we take advantage of blockchain features such as being unchangeable, errorless, distributed, and a single point of failure nonexistence, high security, and proper use in protecting people's privacy. Since a variety of antibody and human health proving tests against the virus are developing, this study tries simultaneously to design an integrated and secure system to meet the authenticity and accuracy of different people's health certificates for the companies requiring these certifications. In this system, on the one hand, there are qualified laboratories that are responsible for performing standard testing and also providing results to the system controller. On the other hand, entities that need to receive health certificates must be members of this system. Finally, people are considered as the end-user of the system. To provide test information for the entities, the mechanism of KYC tokens will be used based on the Stellar private blockchain network. In this mechanism, the user will buy a certain amount of KYC tokens from the system controller. These tokens are charged in the user's wallet, and the user can send these tokens from his wallet to any destination company, to exchange the encrypted health certificate information. Finally, considering the appropriate platform provided by blockchain technology and the requirement of a reliable and accurate solution for issuing health certificates during the Covid-19 pandemic or any other disease, this article offers a solution to meet the requirements.

## 1. Introduction

Today, many strategies are being developed to diagnose the body resistance against the COVID-19 virus. For example, many countries are developing vaccines [1] or antibody diagnostic tests [2]. Ultimately, if any solution will be approved, the request for the safety certificate against Coronavirus will increase dramatically all over the world. Therefore, an attempt has been made to provide a solution for issuing these certificates to deal with this pandemic in such a way that be applicable for any other diseases, whether epidemic or otherwise, like HIV. Based on this issue, an attempt has been made at including the following features in the article proposed solution:

1. Accuracy and irrevocability
2. Except for taking tests, people present in the physical environment is not essential
3. A fast solution
4. The transaction registration and log should be unchangeable to prevent the destructive Interventions of some governments
5. Can be tracked
6. People's privacy should be fully respected
7. Data transfer speed on the secure platform be high
8. Implementing flexibly based on future requirements and not be just a theory.

The Know Your Customer (KYC) refers to a mechanism in which the authenticity of customer information in various fields is validated through a method of verification and authenticity [3]. This validation can confirm financial, identity, or even people's health information. In general, various organizations need to access users' information in various fields. Since the accuracy of the information is very important in areas such as medication, the provided data by users to the units must be reliable. If any unit wants to test and verify the authenticity of the data independently, it will take a long time. Besides, several similar infrastructures will be established by various units to do the same tasks, which will lead to wasting many resources. On the other hand, in each unit, the user has to pay separate financial and time costs to validate his data authenticity. Repeating this process multiple times may cause a difficult user experience.

A reliable, comprehensive system is required to solve the above problems. In this article, the information needed to be validated by the KYC mechanism is people's health testing information of COVID-19 disease. The accuracy of this information is very crucial, and the system can guarantee unchangeable, accessible, and integration of data, based on the mechanisms existing in the blockchain. Following the introduction of blockchain technology in 2008 by Satoshi Nakamoto [4], with features such as immutability, privacy, and transparency, and distribution, blockchain can be used appropriately in the KYC mechanism. Blockchain can be considered a public ledger that stores all transactions in a chain of blocks. In this system, mechanisms such as encryption and various consensus algorithms are used to maintain the validity of transactions and data integrity.

Another feature that can be used in some blockchain networks, such as Ethereum or Stellar, which is used in this article, is the ability to define tokens as digital assets that can be exchanged along with the network's native cryptocurrency [5]. Tokens can be considered a digital representation of anything that can be traded (whether goods or services) issued on a blockchain platform. For example, tokens can be used as a financial or physical asset such as gold or money or as an abstract concept such as an identity certificate. These tokens will be provided to users by the system controller, which verifies the authenticity of the information. By sending these tokens on the blockchain platform, users will be able to allow access to their accurate and guaranteed information to any destination company. In the designed system in this research, these tokens are a scheme related to the abstract hash of people's health certificate, which contains the hash which provides access to the encrypted information of individuals and can be sent to any member of the system.

This solution will enable users to buy a specific number of KYC tokens and sending to various valid destinations after authenticating only one time in the central entity and verifying the authenticity of their health test results. Companies that want to be aware of their customers' health status (hereinafter referred to as destination companies) at the lowest cost have access to accurate data, which will remove all restrictions on traditional systems. Fig.1 shows the problems with traditional KYC systems. The platform to issue the KYC token in this article is the Stellar blockchain network.

Stellar is an open-source, free, and very flexible blockchain platform that allows developers to use it privately and publicly. This platform allows users to easily generate and distribute their favorite tokens with any number on the network. This ability allows the system controller to easily produce as many KYC tokens as it needs on a private network. Each user can purchase as many of these tokens as he needs. Since the blockchain network is private in this article, only the system controller will be able to generate and distribute the tokens. Besides, on the Stellar network, the average time for producing and registering a block is about 5 seconds, which is about 10 minutes for the Bitcoin network [6] and about 12 to 14 seconds for the Ethereum network [7]. On the other hand, due to the nature of consensus algorithms in Stellar, heavy mining processes are not required, and this increases the efficiency and speed of the system to transfer medical information. These capabilities make Stellar a suitable choice to implement the infrastructure needed to send KYC tokens securely and accurately.

According to the mentioned issues, providing a system based on the KYC mechanism using the features available on the Stellar blockchain platform is an effective and efficient solution and helps governments, industries, and people to accurately diagnose the level of immunity to this viral infection. In the rest, this article is divided into the following sections: Section 2 refers to the related works. Section 3 explains the proposed solution, and Section 4 describes a scenario from the user's check-up in the laboratory to the end of the process of presenting results to the destination company. In the last section, the conclusion and future development strategies are presented.

## 2. Research Background

Since this article is a combination of using blockchain in healthcare and KYC, this section investigates several articles

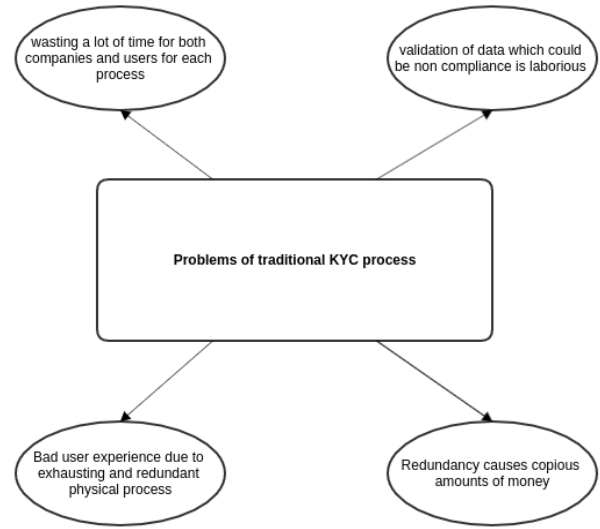

Fig 1. Problems with traditional KYC systems

on blockchain use to store people's medical information and then the use of KYC in the blockchain.

In 2016, Asaph Azaria introduced a system called MedRec, based on which electronic medical records (EMR) data can be stored in a decentralized form with the help of Ethereum's blockchain network. Mechanisms such as information retrieval, access control, data extract method, and the list of medical centers where patients referred (for patients) and the list of patients who have been served (for medical centers) are maintained and controlled through smart contracts on Ethereum platform. The EMR information is located on the local systems of the patient's nodes or the medical center's nodes. This system aims to solve problems such as slow and separate access to patient information, interoperability problems between the system and the Patient Agency, and improving the quality and measure of data for medical research. The incentive mechanism considered in this system, in addition to the intrinsic Ethereum system award, allows access to aggregated and anonymous patient information for data mining purposes [8].

To store people's EMR with the help of the blockchain platform. In 2018, "Fan Kai" introduced a solution called MedBlock to create a more efficient and scalable system. The system includes the following features: A hospital database where users' information is stored. There is also a DLT of blocks containing data such as disease and treatment information, location of finding detailed information, and the Hash of EMR data. There are also nodes in the system that play some roles, which include preparing the primary information, measuring information accuracy, writing the information on the DLT, and removing the problematic nodes. In this system, since each node has only one role, the scalability is greatly increased, and the system becomes more efficient. A significant point in this system is using the Breadcrumbs mechanism, where the encrypted medical information of each person is classified according to the hospital department, which the patient referred to, and searching performs much more efficiently [9].

In 2019, Moyano proposed a solution to optimize the KYC registration process and cut its ever-increasing costs in banks and financial centers. For designing, he used a compilation of the distributed database architecture provided by Siegenthaler and Birman [10]to save information, and the Ethereum blockchain structure to manage and allocate read and write permissions on this distributed database. The process is fulfilled so that once the user goes to one of the system's banks, his information is stored in the database, and a smart contract is created for him, which includes how to access the user's information and the cost of registering KYC. Moreover, if each of the system's banks pays a part of the KYC cost, they will have access to user information [11].

In 2020, Manoj Kumar provided a solution in which, after validating for the first time, the bank creates a key pair and a QR code for the client, which is based on a combination of KYC data, user ID, and user's public key which are available to the user. The information in the QR code is encrypted, and then the user's information is signed, and this signed information is placed inside blockchain along with the public key of the financial entity and the user's ID. After referring to each of the system's banks, the user provides his KYC information to the destination bank using his QR code, and after re-validation through signed data of the previous KYC information in the blockchain, the bank can verify and use information [12]. Table 1 presents a brief comparison of the previous works.

3. THE PROPOSED SOLUTION

The whole presented system in this article is based on how to store and authenticate information using the Stellar blockchain platform. Laboratories and typical users' information, as well as destination companies, is stored in the off-chain section. If the user's check-up result is available and his information is recorded on the off-chain system, he can ask for purchasing a specific number of tokens. These tokens are initially purchased as raw tokens, which contain no information and will be charged to his wallet by the System Controller. By selecting the destination company, which is a specific company using the KYC data, the user receives the hash of his certificate, and after sending it to the destination company, the health information is retrieved through the hash. KYC tokens are used in this system as the essence of registering transactions, and without registering transactions, they have no value or information. This will increase the security of the network and drop the aimless transfer of these tokens.

*3-1. OFF-Chain Section*

The main task of this section is to provide a solution for storing basic system information such as information about users, test results, laboratories, destination companies, and transactions. Moreover, the system is responsible for assigning keys to Entities such as laboratories and destination companies which exchange vital and important user information. It should be noted that the keys allocated in this section are different from the keys in the chain section. Another function of the system is to provide APIs to connect different system components to this section. The duties of this section, along with the main components and the data structure, are shown in Fig.2. The server provides tasks such as exchanging information between components of the system, including users, laboratories, blockchain network, and destination companies. This section registers and modifies or retrieves information in the database if necessary. Logical functions of the system, such as producing hash for transactions, keys for encryption, encryption, or decryption of texts via the RSA method, are also the responsibility of this section. The other part is the database; the schema of the tables can be seen in Fig.2.

In general, in the first step of the off-chain process, the requests are received in the APIs Section, then if the input and the output are correct, the information is transferred to the logic section according to the user's request type. After performing the logical processes, with the help of the DAO section, changes on the data which are caused by the processes, are stored in the database and the appropriate response will be returned.

*3-2. On-Chain Section*

This section describes how to design and implement Stellar blockchain infrastructure to exchange KYC tokens between users and destination companies that need access to accurate personal health information. In general, this part of the system includes a blockchain network to create accounts, generate a pair of blockchain keys, generating and distributing the KYC tokens among users, transferring tokens, and recording transactions. The architecture of this section can be designed as a public or private blockchain network, according to the opinion and discretion of the system controller. Since the exchange of medical data is sensitive and highly confidential, using a private and controllable network by the central company is usually preferred. In this paper, a private blockchain network architecture is supposed to be used.

This platform should also provide the ability to create new tokens and distribute them among network users for the central network administrator in a straightforward and accessible way. In general, this network should be able to provide the following services and features:

1) It can be implemented and used both publicly and privately.
2) Users and destination companies can easily create an account and receive their blockchain public and private keys.
3) Creating a new token by the system controller and distributing it among users easily.
4) Users should be able to send tokens to the desired destinations quickly.
5) The hash of people's health certificate data can be registered in the transactions and fetched by the destination.
6) The mining mechanism for this network is not justified because the transfer must be fast and easy, and the new block must be set up as soon as possible.
7) The token nature is not financial; therefore, tokens must be defined initially in a certain number, and their trade is only the exchange of data on a secure platform, not trading a valuable financial asset.
8) Implementing the system should be fast, easy, and flexible.

Due to these requirements and the native features of the blockchain network mentioned earlier, this network can be a suitable choice for developing the network required by this system.

TABLE 1. COMPARISON OF PREVIOUS WORKS

| Title | Author | Year | Goal |
|---|---|---|---|
| MedRec: Using Blockchain for Medical Data Access and Permission Management | Asaph Azaria, Ariel Ekblaw, Thiago Vieira and Andrew Lippman | 2016 | Creating a decentralized system which uses blockchain platform support for EMR storage |
| MedBlock: Efficient and Secure Medical Data Sharing Via Blockchain | Kai Fan & Shangyang Wang & Yanhui Ren & Hui Li & Yintang Yang | 2018 | Optimizing the EMR distributed storage process on the blockchain and increasing scalability |
| Optimized and Dynamic KYC System Based on Blockchain Technology | José Parra-Moyano, Tryggvi Thoroddsen, Omri Ross | 2019 | Storage and management of bank customers KYC information to prevent multiple data registration and design appropriate incentive mechanism |
| A blockchain-based Approach for an Efficient Secure KYC Process with Data Sovereignty | Dr. Manoj Kumar, Nikhil, Parina Anand | 2020 | Storing KYC information in an encrypted QR code and using blockchain to manage accesses and identify the user while visiting the bank |

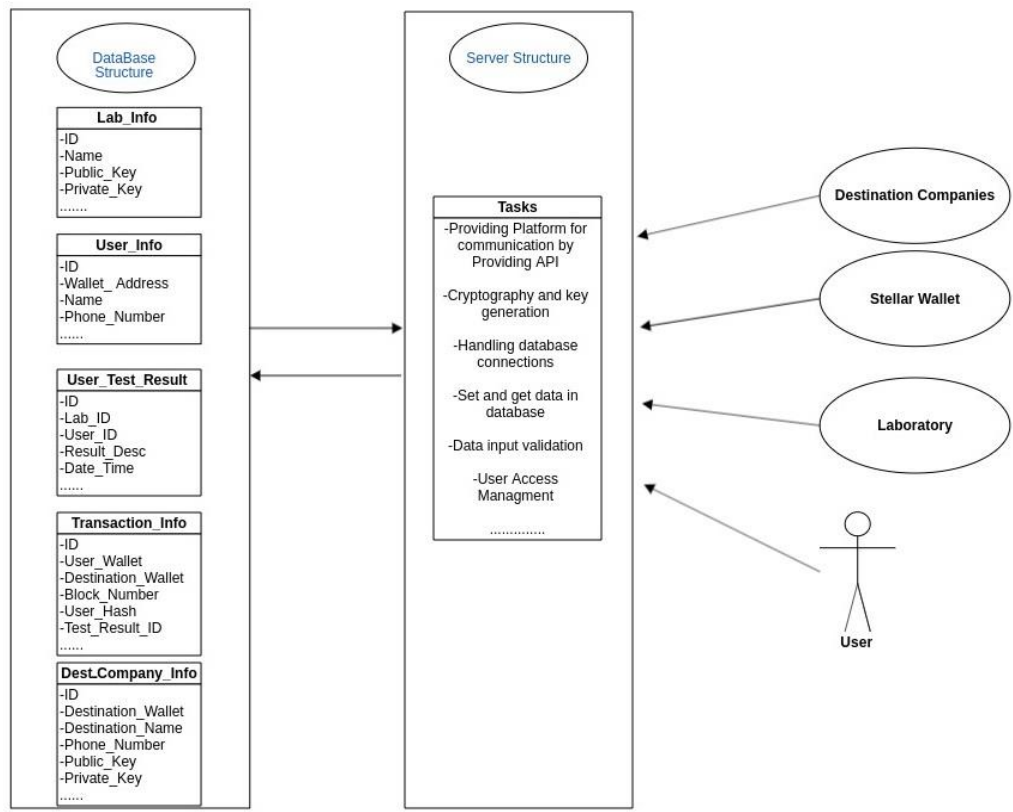

Fig 2. The system main components and data structure

Setting up a network based on the Stellar, which can fully provide all the services in the Stellar system, requires a large amount of money and time because of providing and setting up many servers. This problem Makes independent developers or researchers who cannot prepare these facilities, unable to use and study on this network. Therefore, to solve this problem, Stellar has created a very practical solution. Stellar provides a network of 3 nodes that form a complete blockchain network. This network is generally different from the main Stellar network on which the Lumen cryptocurrency is located. It has also developed a web application for the primary use of network APIs under the A Secure and Efficient Approach for Issuing KYC Token As COVID-19 Health Certificate Based on Stellar Blockchain NetworkHTTP protocol. This application was created to use and test several APIs on the network. Stellar calls this network the Stellar Test Network and the app the Stellar Laboratory [13]. The main reason for developing this laboratory is the use of independent people who usually work on the network for research purposes. This lab provides all the services available on the Stellar network to develop applications. In general, this network is used for the following purposes:

*1) Creating test accounts with test balances for transactions*
*2) Producing and distributing new tokens on a trial basis*
*3) Registering all kinds of transactions from the network sources to any desired destination in the network*
*4) Developing applications and researching and training about Stellar without the possibility of destroying valuable assets*
*5) Testing existing programs*
*6) Performing data analysis on a smaller and insignificant data set compared to the main network.*

In this research, this network has been used to provide transaction services and the KYC tokens in the blockchain network. Since all APIs and network services are the same as the main network, replacing the network with the original version is very easy, both publicly and privately.

*3-3. C.   User Interface*

There will be a user interface that is presented in this section for the actors, including end-users and destination companies to interact with this platform to use system services, and each part will be described separately.

Each user should communicate with the system using an interface and use the services provided in the system. These services include registering the user in the system, verifying the user identity, assigning a unique ID to exchange the tokens, as well as a platform to purchase and send the tokens to the user's required destinations. Some of these services are provided on the blockchain platform, which must be implemented using SDKs provided by Stellar, and some of these are provided privately by the off-chain servers of the system controller. These services must be made available to users and destination companies in an easily consensual accessible platform. In this article, a mobile wallet implemented on the Android operating system will do this task.

In addition to accessing the off-chain services provided by the system controller, this wallet will also include all the services related to the transfer of tokens and authentication at the destination. Although in the current study, this platform has been implemented as a mobile application, it can be completely feasible on various platforms, including a web application.

4. SYSTEM SCENARIO

In this section, the complete system application scenario, along with its design method, is described. In the first step, due to the importance of maintaining the confidentiality of medical and health information of individuals, it is recommended that the blockchain network used in the token exchange subsystem be privately monitored and managed by the system controller. In the first step, after setting up the system, the system controller grants the permission of joining to the approved laboratories which meet the necessary standards in performing their tests. These laboratories will be able to provide services to users if their quality and accuracy are guaranteed by the system controller and are certified to perform correctly. This process can vary depending on the type of test and the requirements of the system controller. After the laboratory membership in the central system, the System Controller assigns a pair of keys and delivers the encryption key to the laboratory. The lab is now ready to provide services to users.

Moreover, destination companies that need to certify the health certificate of individuals registers in this system. At the time of registration, in addition to assigning a pair of cryptographic keys to them, a blockchain account will also be created for these companies. This account will be used by these entities to receive tokens in the future. After registering this information on the server of the System Controller, the keys and information about using the blockchain account will be given to the destination company so that by entering its wallet, it will be able to receive and decrypt the health certificate data.

In the next step, users must register in the system and enter their identity information. After authentication, the user will be shown a list of all valid laboratories in the system. The user selects a laboratory from the list and goes there to test. In the laboratory, to authenticate the data in the future, in addition to the medical test, the user gives biometric information, including fingerprints or any other model of this type of data, if it has not been previously recorded. In the end, after determining the test result, the complete test data, along with the user's biometric information, is encrypted using the encryption key that was originally provided to the lab and sent to the off-chain component of the main server. On the server, this data is decrypted, and the test information is recorded in the database using a unique identifier. Biometric information is also recorded in the user information section. The user will then be able to purchase a KYC token. Fig.3 shows the process of laboratory membership, testing, and recording test results.

To purchase a KYC token, the user must have at least one valid test on the system. By referring to his wallet, the user sends the request to purchase the KYC tokens, up to the maximum amount specified by the system controller. After paying the cost of the tokens, the number of purchasing tokens will be sent to the user's blockchain account, and the user will be able to view the tokens and exchange them in his wallet. These tokens are raw at the time of purchasing and exchange data only at the time of sending with the mechanism that will be explained. Therefore, these tokens do not contain any data before being sent to the destination entity.

To send a token to the destination entity in order to obtain the health certificate, the user must enter the address of the destination entity in his wallet in the token sending section. Consequently, a list of all valid tests for sending to the destinations will be displayed. The user selects the desired test and starts the sending process. On the server-side, the desired test ID, source, and destination of the transaction are received from the user. A 16-byte Hash is then created for this transaction by the off-chain component, which we will

46

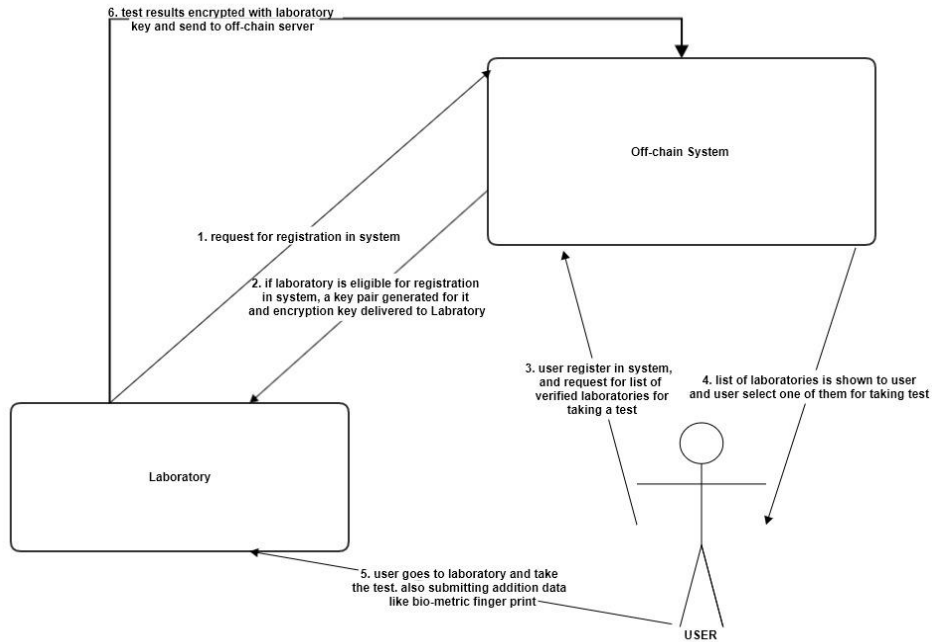

Fig.3. Steps are taken from the registration to the recording result

call it user-hash. The information received from the user is recorded in the transaction database. Further to the above, this database has another field, which is the block-hash in the blockchain. The stellar network generates this hash. This field will become full in the next step.

After successfully registering the transaction information on the server, the created user-hash will be returned to the user wallet. This hash is placed in the Memo section of the transaction block. A transaction from the user's source to the destination company is recorded along with the Memo information on the blockchain network. Fig.4 shows a view of each registered block.

After successful transaction submission, the block-hash information which is received in the wallet will be sent to the server along with user-hash. Block-hash will be added to the transaction data which existed in the server. After this process, the server will respond containing the unique identifier of transaction data.

This identifier is displayed to the user as a numeric code as well as a QR code containing the block-hash, as well as the user is asked to keep this code securely. The user can retrieve the QR code by the numeric identifier. After performing this step, the user has successfully sent his health certificate to the desired destination company. Fig.5 shows a view of this process.

The final step of providing the service by the system is that first, the user attends the destination company and shows his QR Code. If necessary, in addition to the QR code, the physical identity document, such as the passport or biometric information that has already been registered, will be provided to the destination company to start the information fetching operation by them.

For fetching the certificate information, destination companies need to fetch the transaction through the Block-Hash presented as a Qr Code, from the Stellar blockchain platform. Then, through the information in the Memo, this transaction will achieve the test information of the user.

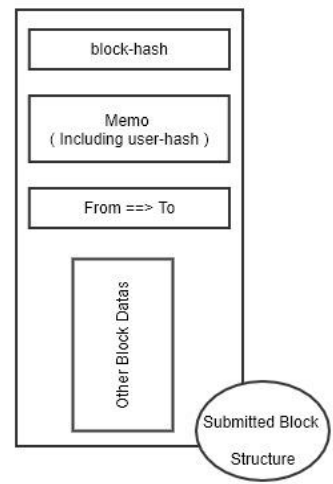

Fig 4 . Registered Block

Access to user information is such that first the hash in the Memo is sent to the Off-chain server. If such a transaction is available on the server, the information of the destination company and the information related to the user's test ID will be fetched, and through this Test ID, the details of the test result will be accessible. Finally, the test result is encrypted with the registered key of the destination company and is returned to it. If the Hash sending company is the same as the destination company registered with the transaction, the sender could decrypt the information using its private key; otherwise, it will not be able to do so.

According to the results of the test, the destination company decides how to provide the services to the user.

In the end, the destination company can resell the collected tokens at a price of one-fifth of the token's original price. In addition to largely preventing the issuance of new tokens,

A Secure and Efficient Approach for Issuing KYC Token As COVID-19 Health Certificate Based on Stellar Blockchain Network

this is seen as a side income for the destination company and encourages it to participate in the suggested system actively.

Fig.6 shows a scheme of the operation from the user attendance moment to the end of the information fetching process.

The solution proposed here has these advantages:

1) *The whole process of the system from testing to the assessment of health certificates is automated.*
2) *Due to using a blockchain network, the proposed system has more reliability than other solutions.*
3) *Business costs will be decreased.*

5. CONCLUSION

Concerning the problem of the COVID-19 pandemic and the many economic and livelihood problems that have arisen, this study provided a solution for issuing health certificates for people, so that, goals such as being reliable, credible, unchangeable, fast, economical, and secure alongside protecting individuals' privacy be achieved.

The proposed system has three sections; off-chain, on-chain, and the user interface. The off-chain tasks contain storing information and registering entities such as destination companies, users, testing results and laboratories, encrypting data, and creating a platform for communication with others were allocated.

In the on-chain, due to the numerous advantages of the Stellar blockchain network and the compatibility of this platform with the basic needs in this article, a Stellar blockchain network was designed and implemented using the Stellar test network infrastructure. In this strategy, the network can be developed both publicly and privately. Still, due to the nature of the medical data sensitivity, the private type is recommended for use in the operating environment. This network was used to produce raw KYC tokens as well as secure and unchanged data exchange to purchase and exchange tokens from users to destination companies that need medical data. Due to blockchain features, this platform will guarantee the authenticity of the data and will provide fast access to valid data at a lower cost. In addition, to use the services of this comprehensive system, a user interface is considered as an important part of the system. In this article, an Android mobile wallet was designed as the user interface. This wallet is the consensus of all existing services for users. Token destination companies can also use this wallet to authenticate the received tokens.

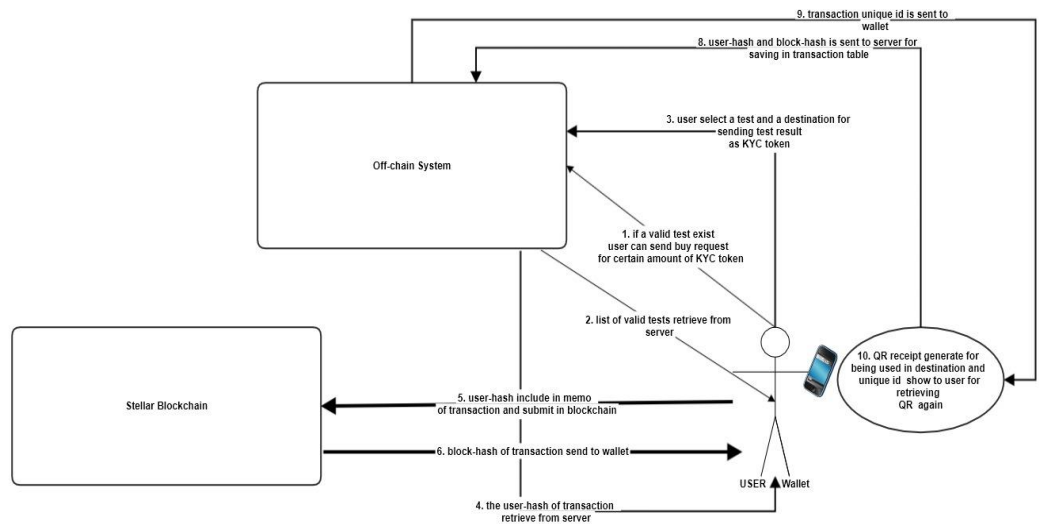

Fig 5. A scheme of the transaction registration

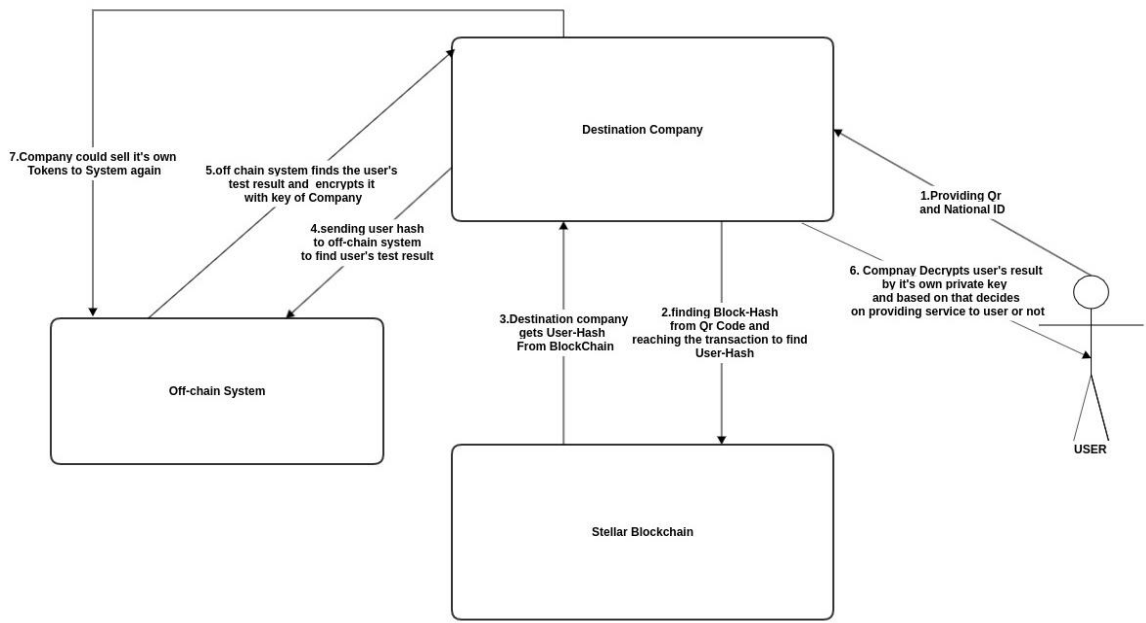

Fig 6. The process of providing information to the company and receiving services

Finally, in addition to the benefits mentioned above, providing a comprehensive archive for laboratory clients to be needless of launching local archives is an advantage for the system. Moreover, the resale of tokens received by destination companies to the main system and the creation of side income for those companies are also among the most important incentives and benefits of the system.

The use of such systems can increase the readiness of the human race to cope with the COVID-19 pandemic or any other problem in the future that requires the approval of human health so that no great losses will be seen anymore.

Our immediate goal for the future is to prepare a big data platform for using aggregated data in the off-chain section. The final purpose is to provide a comprehensive data analysis report for governments and all official health centers to help them in planning and decision making for society.

## References

[1] T. T. Le, Z. Andreadakis, A. Kumar, R. G. Román, S. Tollefsen, M. Saville, and S. Mayhew, "COVID-19 therapies and vaccine landscape", Nature Materials, vol. 19, no. 809, pp.305-306 , 2020.

[2] A. Petherick, "Developing antibody tests for SARS-CoV-2", The Lancet, vol. 395, no. 10230, pp. 1101–1102, 2020.

[3] N. Kapsoulis, A. Psychas, G. Palaiokrassas, A. Marinakis, A. Litke, and T. Varvarigou, "Know Your Customer (KYC) Implementation with Smart Contracts on a Privacy-Oriented Decentralized Architecture", Future Internet, vol. 12, no. 2, p. 41, 2020.

[4] C. S. Wright, "Bitcoin: A Peer-to-Peer Electronic Cash System", SSRN Electronic Journal, 2008. Available at https://ssrn.com/abstract=3440802

[5] M. Borkowski, D. McDonald, C. Ritzer and S. Schulte, "Towards Atomic Cross-Chain Token Transfers: State of the Art and Open Questions Within TAST", Distributed Systems Group TU Wien (Technische Universit at Wien), 2019, [online] Available: https://dsg.tuwien.ac.at/projects/tast/pub/tast-white-paper-1.pdf

[6] E. Erdin, M. Cebe, K. Akkaya, S. Solak, E. Bulut, and S. Uluagac, "A Bitcoin payment network with reduced transaction fees and confirmation times", Computer Networks, vol. 172, p. 107098, 2020.

[7] L. Zhang, B. Lee, Y. Ye, and Y. Qiao, "Ethereum Transaction Performance Evaluation Using Test-Nets", in Euro-Par 2019: Parallel Processing Workshops Lecture Notes in Computer Science, Springer, Aug. 2019, pp. 179–190,.

[8] A. Azaria, A. Ekblaw, T. Vieira, and A. Lippman, "MedRec: Using Blockchain for Medical Data Access and Permission Management", 2016 2nd International Conference on Open and Big Data (OBD), IEEE, 2016, pp. 25–30.

[9] K. Fan, S. Wang, Y. Ren, H. Li, and Y. Yang, "MedBlock: Efficient and Secure Medical Data Sharing Via Blockchain", Journal of Medical Systems, vol. 42, no. 8, p. 136, 2018.

[10] M. Siegenthaler and K. Birman, "Sharing Private Information Across Distributed Databases:", 2009 Eighth IEEE International Symposium on Network Computing and Applications, IEEE, 2009, pp. 82–89.

[11] J. P. Moyano, T. Thoroddsen, and O. Ross, "Optimised and dynamic KYC system based on blockchain technology", International Journal of Blockchains and Cryptocurrencies, vol. 1, no. 1, pp. 85–106, 2019.

[12] M. Kumar, Nikhil, P. Anand, "A Blockchain Based Approach For An Efficient Secure KYC Process With Data Sovereignty", International Journal of Scientific & Technology Research, vol. 9, no. 1, pp. 3403-3407, January, 2020.

[13] S. Team, "Stellar Laboratory", 2020. [Online]. Available: https://laboratory.stellar.org/. [Accessed: 29-Jul-2020].

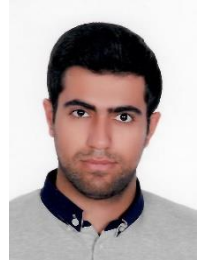

**Kiarash Shamsi** has been received the master of software engineering from the department of computer engineering at the University of Science and Culture, Tehran, Iran. His research interests include Blockchain, data science and distributed systems.

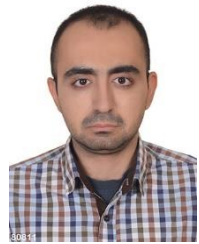

**Koosha Esmaeilzadeh Khorasani** has been received the bachelor of hardware engineering from the department of computer engineering at the Iran University of Science and Technology, Tehran, Iran. His research interests include machine learning, blockchain, and IOT.

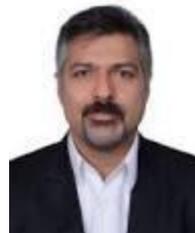

**Mohammad Javad Shayegan** is an Associate Professor at the Department of Computer Engineering at the University of Science and Culture, Tehran, Iran. He is the founder of the web research center, International Conference on Web Research, and International Journal of Web Research in Iran. His research interests include Web research, data science and distributed systems.